\documentclass{article}
\usepackage{spconf,amsmath,graphicx,hyperref}
\usepackage{cite}
\usepackage{amsmath,amssymb,amsfonts}
\usepackage{algorithmic}
\usepackage{textcomp}
\usepackage{xcolor}
\usepackage[table]{xcolor}
\usepackage{multirow}
\usepackage{booktabs}
\usepackage{cleveref}
\usepackage{bigdelim} 

\title{When Audio Generators Become Good Listeners: Generative Features for Understanding Tasks}
\name{\parbox{\linewidth}{\centering
  Zeyu Xie$^{1,2}$, Chenxing Li$^{2}$, Xuenan Xu$^{3}$, Mengyue Wu$^{3}$, Wenfu Wang$^{2}$,\\
  Ruibo Fu$^{2}$, Meng Yu$^{2}$, Dong Yu$^{2}$\textsuperscript{*}, Yuexian Zou$^{1}$\textsuperscript{*}\thanks{* Corresponding authors}
}}

\address{$^{1}$ Guangdong Provincial Key Laboratory of Ultra High Definition Immersive Media Technology, \\
Peking University, Shenzhen  \\
       $^{2}$ Tencent AI Lab, Seattle $^{3}$ Shanghai Jiao Tong University, Shanghai \\
\textit{zeyuxie25@stu.pku.edu.cn}, \textit{zouyx@pku.edu.cn}
}

\begin{document}
%\ninept
%
\maketitle
\begin{abstract}
This work pioneers the utilization of generative features in enhancing audio understanding.  
Unlike conventional discriminative features that directly optimize posterior and thus emphasize semantic abstraction while losing fine-grained details, audio generation models inherently encode both spatiotemporal perception (capturing local acoustic texture across time and frequency) and semantic prior (knowing what to generate).
It motivates us to explore the bridge of these complementary strengths. 
We provide a systematic investigation of their differences and complementary relationships, and  ultimately propose an effective fusion strategy. 
Experiments across multiple tasks, including sound event classification, tagging, and particularly the fine-grained task of audio captioning, demonstrate consistent performance gains. 
Beyond empirical improvements, this work more importantly introduces a new perspective on audio representation learning, highlighting that generative–discriminative complementarity can provide both detailed perception and semantic awareness for audio understanding.
\end{abstract}

\section{Introduction}
\label{sec:intro}

\begin{figure}[tbp]
\centerline{\includegraphics[width=1.0\linewidth]{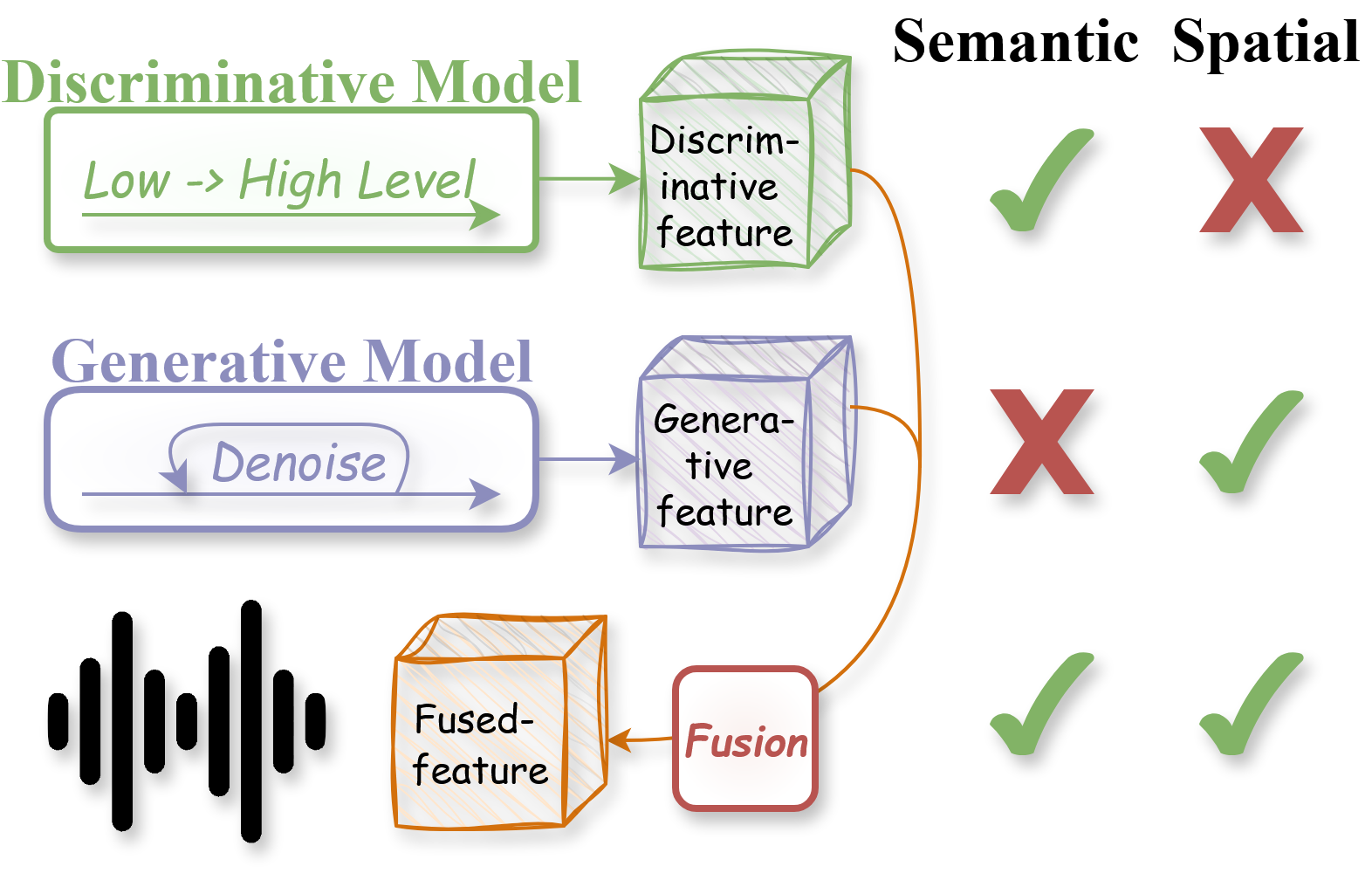}}
\caption{
%Properties of different features.
The spatial information in \textcolor{purple!40!blue}{generative features} complements the semantic information in \textcolor{green!80!black!}{discriminative features}.
}
\label{fig:sample}
\end{figure}

Current audio understanding models utilize stacked neural networks to learn and classify acoustic features, which involves aggregating knowledge from sparse acoustic features to compact semantic features.  
However, this abstraction process discards acoustic details, such as the spatio-temporal characteristics of sound.
%Simply put, this involves aggregating knowledge from sparse acoustic features to compact semantic features. 
%However, this process ignores acoustic details, including the temporal and spatial characteristics of sound. 
Audio generation is an opposite process, where the generated features need to have high-quality and sufficient spatiotemporal details.
Hence, generative and discriminative features exhibit intrinsic complementarity, as shown in Figure~\ref{fig:sample}.
Motivated by this, we innovatively introduce generative features to enhance audio understanding.
%Building on this observation, we innovatively introduce generative features to enhance the perception of fine-grained details in audio understanding tasks.
% As shown in Figrue~\ref{fig:sample}
% we demonstrate that generative and discriminative features are intrinsically complementary to some extent, a conclusion supported by empirical visual comparisons in Figure~\ref{fig:cmp}.
% Exploring how to use audio generation models to generate useful representations and effectively incorporate them into audio understanding tasks has practical and theoretical significance.

%Relatedwork: Audio understanding Model
% HTS-AT~\cite{} introduces a hierarchical Transformer structure inspired by Swin-Transformer~\cite{}, reducing model size and training costs while supporting audio event localization through token-semantic mapping.
% BEATs~\cite{} leverages a self-supervised framework with joint optimization of an acoustic tokenizer and Transformer, employing discrete label prediction instead of reconstruction to capture richer semantics.
% Dasheng~\cite{}, based on a masked autoencoder, is trained on massive diverse audio data to bridge the generalization gap across speech, music, and environmental domains.
% EAT~\cite{} improves efficiency and effectiveness through bootstrap training and large-block masking, achieving state-of-the-art performance with significantly faster pre-training.
% These models typically preprocess audio by chunking or patching the input signal, enabling effective representation learning within the Transformer framework.
\textbf{Discriminative audio representations are semantically rich yet spatially deficient features}.
Mainstream audio understanding models utilize discriminative features characterized by high-level semantic information. 
%PANNs~\cite{kong2020panns},
For instance, in the audio tagging (AT) task, supervised models like  AST~\cite{gong2021ast}, and HTS-AT~\cite{chen2022hts} use event labels as supervision for semantic extraction;
Self-supervised models BEATs~\cite{chen2022beats}, Dasheng~\cite{dinkel2024scaling}, and EAT~\cite{chen2024eat} employ masking strategies to learn semantic representations.
%are trained by leveraging event labels as supervisory signals for semantic extraction;
% Among self-supervised learning (SSL) models, BEATs~\cite{chen2022beats}, Dasheng~\cite{dinkel2024scaling}, and EAT~\cite{chen2024eat} use masked data and specialized training methods to learn rich semantic representations.
These semantically-rich features are also widely utilized in transfer learning.
Xu et al.~\cite{xu2021investigating} and ACT~\cite{mei2021audio} use a pretrained audio encoder to enhance audio caption generation.
LAION-CLAP~\cite{laionclap2023} employs the HTS-AT to extract audio semantic features for contrastive learning.
%alignment with text embeddings.
% Furthermore, LAION-CLAP~\cite{laionclap2023} employs the HTS-AT model to extract audio semantic features for contrastive learning alignment with text embeddings.
Regardless of the specific methodology, the discriminative features utilized by these models are designed to abstract specific compact semantics (e.g., class-discriminative feature in AT) from sparse audio data. 
Their feedforward architectures leverage the hierarchical depth of the network, to progressively extract semantic representations from low-level details through layer-wise abstraction.
This results in a loss of fine-grained audio details.

%Relatedwork: Audio generation Models
\textbf{In contrast, generative features exhibit rich spatial characteristics while retaining semantic information.}
As text-to-audio generation model can effectively convert text into audio, it possesses the ability to perceive both the identity and location of events in audio clip.
This perception is achieved through interactions between embedded text and internal representations in generative modules, indicating that these internal representations encode both location of events and correlations with semantic concepts from text.
%event layout concepts and correlations with semantic concepts from text.
% descriptions

% Recent text-to-audio (TTA) generation models—such as AudioLDM2~\cite{liu2023audioldm2}, Tango2~\cite{majumder2024tango}, and EZAudio~\cite{hai2024ezaudio}—can generate highly realistic audio from text prompts. 
% Recent text-to-audio (TTA) generation models have demonstrated the ability to produce highly realistic audio from textual prompts~\cite{liu2023audioldm2,majumder2024tango,hai2024ezaudio}.
% These models produce high-quality sound events with strong semantic alignment. 
Recent text-to-audio (TTA) models can generate highly realistic and semantically aligned audio from text prompts~\cite{liu2023audioldm2,majumder2024tango}.
%thereby generating sound events with strong semantic alignment.
% models have demonstrated the ability to produce highly realistic audio from textual prompts~\cite{liu2023audioldm2,majumder2024tango,hai2024ezaudio}, thereby generating high-quality sound events with strong semantic alignment.
Motivated by the aforementioned observations, we propose a hypothesis:
\textbf{The innate feature extraction capabilities of advanced generative models could be effectively leveraged to enhance audio understanding tasks}.

%Motivated by the aforementioned observations, we pose the following research question: \textbf{Can the innate feature extraction capabilities of audio generative models be effectively leveraged in audio understanding tasks?}

%Relatedwork: Some models combine audio understanding and generation.
There is currently no research on how to leverage the unique advantages of generative feature for enhancing audio understanding models.
Although prior studies like DiffATR~\cite{xin2024diffatr} have employed generative models in audio understanding, they focus on joint multimodal distributions rather than feature utilization.
%Although some studies have explored the use of generative models for understanding tasks — such as DiffATR~\cite{xin2024diffatr} — they primarily focus on the joint distribution of the two modalities rather than on information extraction.
This work analyzes properties and distinctions of generative versus discriminative features, and demonstrates the ability of generative features to provide complementary information in downstream tasks, particularly demonstrating significant performance gains on the fine-grained task of audio captioning.
Our contributions are:
\begin{enumerate}

\item To the best of our knowledge, we introduce generative features for audio understanding tasks for the first time.

% \item 
% We introduce the first spatiotemporally-aligned audio generation model, which is subsequently utilized to extract generative features.

\item 
We present the first analysis of generative features derived from audio effect processing, comparing them with discriminative features.

\item 
We validate our approach on downstream tasks, demonstrating that fine-grained understanding tasks benefit from the incorporation of generative features.

\end{enumerate}

\section{Generative Features}
\label{sec:gen_feat}

Current TTA generation models typically follow a modular design: (1) a Variational Autoencoder (VAE) compresses raw audio into compact latent representation, (2) a text encoder extracts meaningful semantic embeddings from text prompts, and (3) a generative module predicts the latent embedding conditioned on the textual semantics.
The intermediate latents from the generative module can be utilized as generative features.
To effectively leverage those features, it is essential to first investigate their \underline{fundamental characteristics}.

\underline{Denoising}: The core of the generation process leverages a denoising objective: the model first corrupts the audio latent representation by introducing noise and is subsequently trained to reconstruct the original latent from the noisy input. 
During inference, it iteratively recovers the audio from noise.
This strategy underpins many recent diffusion-based and flow-based generative approaches~\cite{liu2023audioldm2,majumder2024tango}.
Therefore, such generative models inherently possess a denoising capability, enabling them to remove estimated noise from inputs.
%Therefore, such generative models inherently possess a denoising capability, allowing them to \textbf{remove estimated noise} from a given input. 

\underline{Semantic Extraction}: During the iterative denoising process, the model computes attention weights based on the conditional text embedding and the current denoised audio latent. %audio latent representations 
This enables the intermediate hidden representations of the model to exhibit high-level features correlated with the textual input—that is, semantically meaningful information.
%\textbf{semantically meaningful information}.

\begin{figure}[tb]
\centerline{\includegraphics[width=1.0\linewidth]{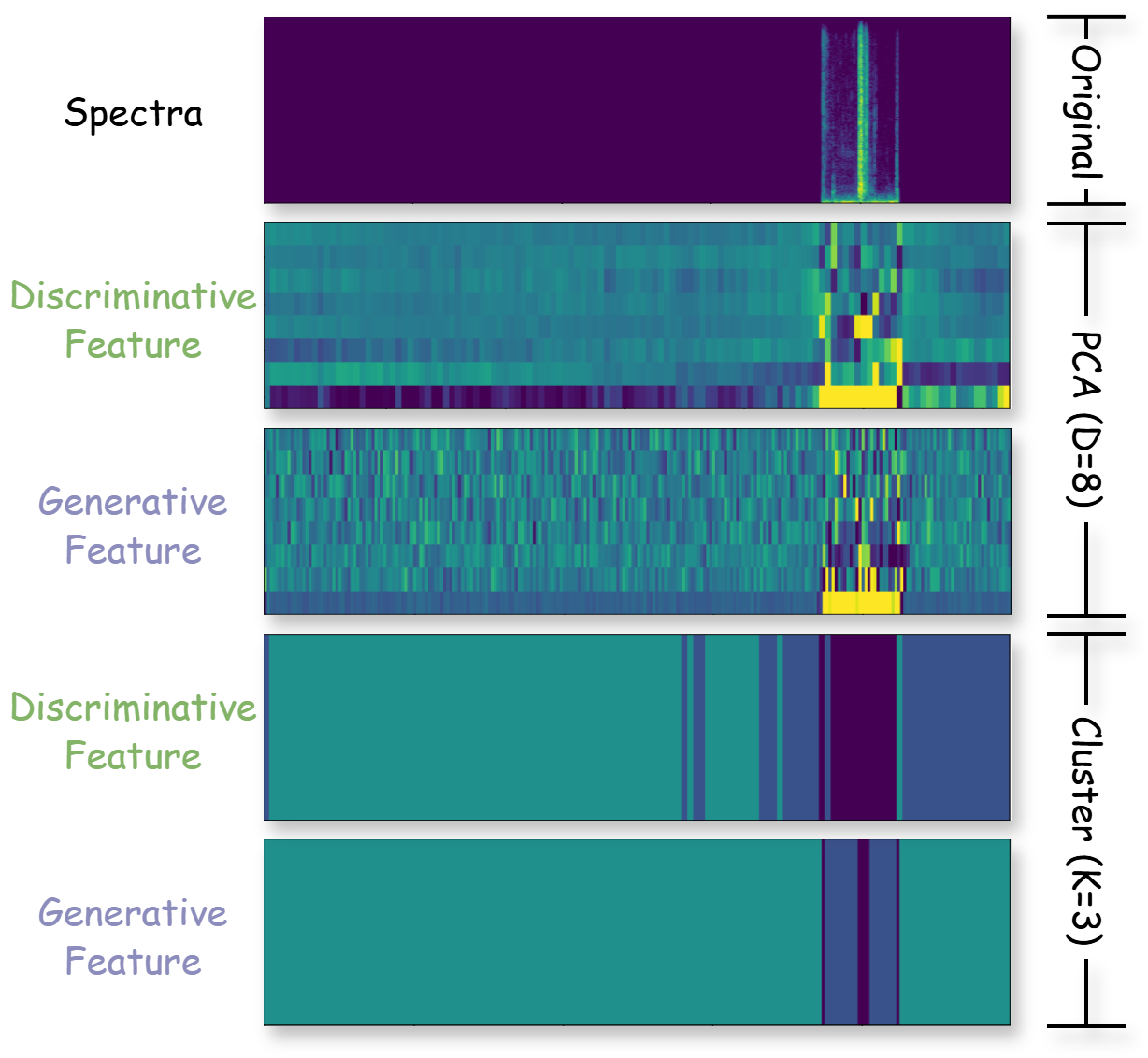}}
\caption{
%Comparison between Mel features, discriminative features, and generative features. 
Visualization of features.
%\textcolor{purple!40!blue}{Generative features}  exhibit more distinct boundaries and higher sensitivity to the structural contours of sound events than \textcolor{green!80!black!}{discriminative features}.
\textcolor{purple!40!blue}{Generative features} exhibit clearer spectral structures and more defined temporal contours compared to \textcolor{green!80!black!}{discriminative features}.
}
\label{fig:cmp}
\end{figure}

%\underline{Spatiotemporality}: Most importantly, the generative models exhibit \textbf{strong sensitivity to the spatiotemporal characteristics} of audio. 
\underline{Spatiotemporality}: Most importantly, the generative models exhibit sensitivity to the spatiotemporal characteristics of audio. 
Throughout the generation process, the model progressively refines event localization, enhances structural contours, and optimizes fine-grained details in a coarse-to-fine manner. % across denoising stages.
In the final steps, it further sharpens the textural details of auditory events, thereby producing clearly defined perceptual contours and rich spatial information, as shown in Figure~\ref{fig:cmp}.
% Throughout the generation process, the model progressively refines event localization, enhances structural contours, and ultimately optimizes fine-grained details. 
% It improves the perception of event positions in a coarse-to-fine manner across the denoising stages. 
%Especially in the final steps, i
% It refines the detailed textures of auditory events, thereby producing clear perceptual contours and rich spatial information, as indicated in the bottom row of Figure~\ref{fig:cmp}.
%This hierarchical generative process explains its ability to form coherent auditory structures and precise temporal boundaries. 
%A conceptual illustration of this mechanism is provided in Figure 2.

Hence generative models are inherently suitable for (1) effectively capturing spatial information, (2) using as semantic feature extractors while (3) offering simultaneous denoising.
% Consequently, generative models are inherently (1) suitable for use as audio semantic feature extractors, (2) offering simultaneous denoising capability that removes interference while (3) effectively capturing spatial information.
% Most importantly, the generative model progressively reconstructs structural and spatial information throughout the denoising stages. 

\begin{figure*}[tbp]
\centerline{\includegraphics[width=1.0\linewidth]{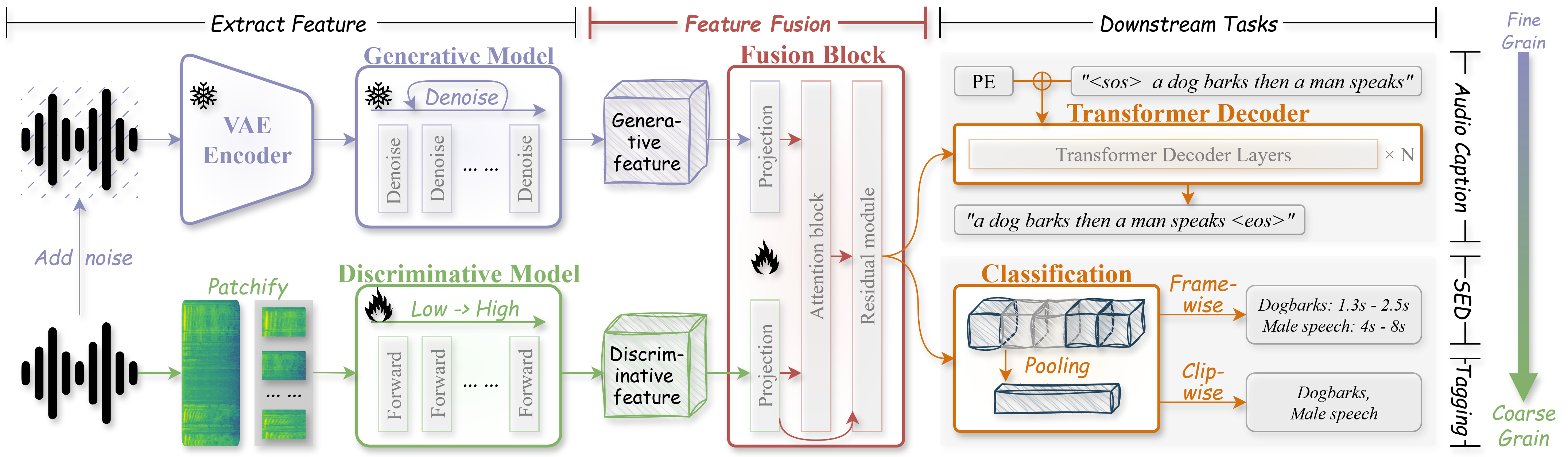}}
\caption{
Illustration of integrating generative features into understanding tasks. 
Both generative and discriminative features are extracted by their respective models, processed through a fusion module, and subsequently delivered to downstream tasks.
}
%An illustration of .}
\label{fig:pipeline}
\end{figure*}

\section{Comparison between generative and discriminative features}
% Mainstream audio understanding models predominantly utilize discriminative features. 
% These features, such as those derived from supervised or self-supervised pre-trained models, capture rich semantic information because they are trained to identify and emphasize perceptually salient patterns from audio. 
% As a result, they demonstrate strong performance in downstream tasks like audio tagging.

% However, the high-dimensional feature extraction process in these models often leads to a partial loss of spatiotemporal information. 
% This limitation becomes apparent in tasks requiring fine-grained temporal or structural details.
% From a modeling perspective, grouping features into patches can disrupt the original structural organization of audio, as the local continuity between feature vectors is not entirely preserved.
% Moreover, since discriminative models are not explicitly trained to remove noise, the features they extract may retain irrelevant information and interference, introducing potential errors.
%A defining property of discriminative features lies in their use of deep, hierarchical abstraction within neural networks. 
%This process maps sparse audio representations into a high-level compact semantic representations (e.g., for representing sound event classes).
A defining property of discriminative features lies in their use of hierarchical abstraction in deep networks for information distillation. %to distill information.
They maps sparse, low-level audio representations into compact, high-level semantic embeddings (e.g., for representing event classes). 
However, with increasing depth, networks discard fine acoustic details—each layer amplifies dominant patterns while filtering out subtler information.
% As the network depth increases, the feature becomes more abstract and semantically enriched; 
%However, it also leads to the loss of fine-grained acoustic details, as each layer prioritizes salient patterns while suppressing peripheral or high-frequency information.

In contrast, as described in Section~\ref{sec:gen_feat}, the generative features retain rich spatio-temporal information.
We further applied Principal Component Analysis (PCA, dimension=8) for dimensionality reduction and K-means (K=3) for clustering to compare these two features, as shown in Figure~\ref{fig:cmp}.
The PCA results demonstrate that generative features exhibit clearer spectral structures, while discriminative features show blurred transitions at boundary regions. 
The clustering results also reveal that generative features possess more defined temporal contours. 
This indicates the superiority of generative features in preserving spatio-temporal characteristics.

Discriminative features (rich in semantic information) and generative features (preserving spatiotemporal structure) are complementary to some extent. 
Therefore, we hypothesize that \textbf{integrating these two feature types may yield more fine-grained and informative audio representations}.
%compared to discriminative features.
% The PCA results demonstrates that the generative features exhibit more distinct boundaries. 
% Furthermore, K-means clustering experiments indicate that these features demonstrate higher sensitivity to the structural contours of sound events.

% Discriminative features—which are rich in semantic information—and generative features—which inherently denoise and preserve spatiotemporal structure, are complementary to some extent. 
%Therefore, we hypothesize that the integration of both feature types may lead to more robust and informative audio representations.

\begin{table*}[htbp]
    \centering
    \caption{
    Experimental result on AudioCaps of fine-grain task AAC. 
    SPIDEr is computed as the average of CIDEr and SPICE. 
    ``Params" refer to the number of trainable parameters.
    The fusion models are designed based on the ACT-M architecture.
    }
    \label{tab:audiocaps}
    
    \centering
    \begin{tabular}{l|c|cccccccc|c}
        \toprule
        %\midrule
        \multicolumn{1}{c|}{\textbf{System}} & \multicolumn{1}{c|}{Params} & $\text{BLEU}_1$ & $\text{BLEU}_2$ & $\text{BLEU}_3$ & $\text{BLEU}_4$ &  METEOR & $\text{ROUGE}_\text{L}$ & CIDEr &
    SPICE & SPIDEr\\
    
        \midrule
        ACT-M & 108.01 &0.653 & 0.495 &0.363 &0.259 &0.222 &0.471 &0.663 &0.163 & 0.413\\
        ACT-L & 116.42 &0.647 & 0.488 &0.356 &0.252 &0.222 &0.468 &0.679 &0.160 & 0.420\\
        \midrule 
        % replace &108.15 &0.631 &0.461 &0.333 &0.242 &0.210 &0.457 &0.576 &0.149 &0.362 \\
        % early fusion &108.03  &0.676 &0.501 &0.356 &0.245 &0.227 &0.483	&0.664 &0.164 &0.414 \\
        % mid fusion &108.62 &\textbf{0.685} &\textbf{0.513} &\textbf{0.375}	&\textbf{0.270}	&\textbf{0.236}	&\textbf{0.484}	&\textbf{0.696}	&\textbf{0.169} &\textbf{0.432}\\
  		
        Replace& 108.31  &\textcolor{gray!70}{0.558} &\textcolor{gray!70}{0.380} &\textcolor{gray!70}{0.254} &\textcolor{gray!70}{0.161} &\textcolor{gray!70}{0.164} &\textcolor{gray!70}{0.387} &\textcolor{gray!70}{0.342} &\textcolor{gray!70}{0.107} &0.225 \\
        
         Early fusion& 108.04 &\textcolor{gray!70}{0.684} &\textcolor{gray!70}{0.515} &\textcolor{gray!70}{0.372} &\textcolor{gray!70}{0.263} &\textcolor{gray!70}{0.233} &\textcolor{gray!70}{0.486} &\textcolor{gray!70}{0.683} &\textcolor{gray!70}{0.168} &0.426 \\
        
         %reverse fusion& 109.13 &\textcolor{gray!70}{0.685} &\textcolor{gray!70}{0.512} &\textcolor{gray!70}{0.374} &\textcolor{gray!70}{0.268} &\textcolor{gray!70}{0.234} &\textcolor{gray!70}{0.483} &\textcolor{gray!70}{0.689} &\textcolor{gray!70}{0.170 }&0.429 \\
        
          \rowcolor{gray!20} Mid fusion &108.67 &\textbf{0.693}  &\textbf{0.527} &\textbf{0.387} &\textbf{0.277} &\textbf{0.234} &\textbf{0.492} &\textbf{0.708} &\textbf{0.164} &\textbf{0.436}\\

        %&  & & & & & & & & & \\
        % &  & & & & & & & & & \\

        %\midrule
        \bottomrule
    \end{tabular}
    
\end{table*}

% \begin{table}[htbp]
%     \centering
%     \caption{Cross Domain. Experimental result.}
%     \label{tab:clotho}
    
%     \centering
%     \begin{tabular}{c|cccc}
%         \toprule
%         \midrule
%         \multirow{1}{*}{\textbf{System}} & $\text{BLEU}_4$ &  METEOR & $\text{ROUGE}_\text{L}$ & SPIDEr\\
    
%         \midrule
%         ACT m  &0.135 &\textbf{0.165} &\textbf{0.359} &0.226\\
        
%         mid fusion  &\textbf{0.136} &\textbf{0.165} &\textbf{0.359} &\textbf{0.229}\\

%         \midrule
%         \bottomrule
%     \end{tabular}
    
% \end{table}

\begin{table}[htbp]
    \centering
    \caption{Coarse-grain  AT \& SED.
    ``ft" indicates fine-tuning. }
    \label{tab:audioset_desed}
    
    \centering
    \begin{tabular}{c|c|c}
        \toprule
        %\midrule
        \multirow{2}{*}{\textbf{System}} &Audio Tagging & Sound Event Detection \\
        \cline{2-3}
         & mAP & Macro-F1\\
    
        \midrule
        HTS-AT  &41.6 &39.7\\
        Replace-ft &22.4 &22.2 \\
        Eearly-ft &40.7 & 40.1\\
        \rowcolor{gray!20} Mid-ft &\textbf{41.7} &\textbf{40.7} \\

        %\midrule
        \bottomrule
    \end{tabular}
    
\end{table}

\section{Experiments}
% We validate the efficacy of incorporating generative features across a range of downstream tasks. 
% These tasks are organized, in descending order of their requirement for fine-grained detail, as follows: Automated Audio Captioning (AAC), which involves generating free-text descriptions of audio content; Sound Event Detection (SED), which entails classifying audio events and temporally localizing their occurrences; and Audio Tagging (AT), which focuses solely on the classification of audio events. 
%To verify our hypothesis, we comprehensively validate the efficacy of incorporating generative features across various downstream tasks from multiple perspectives.
%We comprehensively validate the efficacy of incorporating generative features across a variety of downstream tasks from multiple perspectives. 
%Based on the varying levels of detail required, we conduct experiments on $3$ distinct tasks:   (2) Sound Event Detection (SED), which entails classifying audio events and temporally localizing their occurrences; and (3) Audio Tagging (AT), which focuses solely on the coarse-grained classification of audio events without requiring temporal or descriptive detail.
%Based on the varying granularity of feature requirements—progressing from fine to coarse—we evaluate feature performance across $3$ downstream tasks:
To verify our hypothesis, we evaluate the integration of generative features on $3$ downstream tasks with varying granularity requirements—progressively from fine to coarse:
(1) AAC, which generates text descriptions of audio content, requiring detailed depictions;
%of event sequences, attributes, and quantitative characteristics;
(2) Sound Event Detection (SED), which classifies sound events and localizes their temporal occurrences;
(3) AT, which focuses exclusively on coarse-grained classification of events without temporal or descriptive detail.
%These tasks represent a progressive shift from fine- to coarse-grained feature requirements: AAC not only necessitates the description of sound events but also demands a comprehensive summary of their sequential, attributive, and quantitative characteristics; SED requires the accurate identification and timing of individual events; whereas AT is concerned exclusively with classification.

\subsection{Baseline Model}
ACT~\cite{mei2021audio} is adopted as the baseline model for AAC.
It has a Transformer-based architecture that captures global context and temporal relationships in audio via attention mechanisms.
The ACT encoder is pre-trained on AudioSet~\cite{gemmeke2017audio} with AT task to learn more semantic representations.
%We employ the ACT~\cite{mei2021audio} as our baseline model for audio captioning. 
%It implements a fully convolution-free encoder-decoder architecture based exclusively on Transformer networks. 
%It utilizes self-attention mechanisms to capture global contextual information and temporal relationships within audio signals. 
%Specifically, ACT leverages the DeiT model pre-trained on image classification tasks to initialize the parameters of its encoder. 
%Furthermore, the ACT encoder undergoes additional pre-training on AudioSet~\cite{gemmeke2017audio} using an audio tagging objective to learn more semantic audio representations.

HTS-AT~\cite{chen2022hts} is employed as the baseline model for both SED and AT tasks, owing to its dual capability of predicting both frame-wise and clip-level outputs. 
HTS-AT introduces a hierarchical Transformer audio encoder to organize audio inputs, which enables multi-scale capturing of semantic information across different abstraction levels.
%into hierarchical token representations.
% This design enables multi-scale processing of audio information across different abstraction levels. 
By incorporating a token-semantic convolutional neural network (CNN), it maps the final hierarchical representations to class-specific feature maps, thereby supporting not only audio classification but also temporal localization for event detection.

\subsection{Feature Extractor \& Fusion}
TANGO-2~\cite{majumder2024tango} serves as an effective audio generation model and is consequently adopted as feature extractor. 
For an input audio, it performs a 1-step noising followed by a 1-step denoising with an empty text prompt.
The latent from the final layer (before the VAE) is utilized as generative features.
%as generative features.

%Both baseline models consist of an audio encoder followed by other task-specific modules.
%a decoder or a classification module.
Both baseline models have an audio encoder.
We investigate several feature fusion strategies:
(1) Replacement: entirely substituting discriminative features;
(2) Early / (3) Mid-encoder fusion: applying cross-attention before / after the encoder.
%(4) Reverse fusion: integrating generative features as queries after encoder.
Figure~\ref{fig:pipeline} specifically illustrates the mid fusion strategy.
% A natural approach is to fuse discriminative and generative features at this stage, referred to as mid-fusion. 
% As shown in Figure~\ref{fig:pipeline}, We implement an attention module to integrate the discriminative features from the encoder with the externally injected generative features. 
% A residual connection is further incorporated to preserve the original semantic information of the discriminative features throughout the fusion process.

\subsection{Experimental Setup}
All baseline models and training strategies are implemented following the official releases\footnote{ACT: \href{https://github.com/XinhaoMei/ACT}{\textcolor{cyan}{\textit{https://github.com/XinhaoMei/ACT}}}, \\  HTS-AT:  
\href{https://github.com/RetroCirce/HTS-Audio-Transformer}{\textcolor{cyan}{\textit{https://github.com/RetroCirce/HTS-Audio-Transformer}}}
}.
%\url{github.com/RetroCirce/HTS-Audio-Transformer}}.

ACT uses a configuration with $12$ transformer encoder layers and $4$ / $6$ decoder layers in its middle (M) / large (L) settings.
It is trained on AudioCaps~\cite{kim2019audiocaps} using the Adam optimizer.
% The input consists of log-mel spectrogram patches. 
% The model is trained on the AudioCaps~\cite{} dataset using the Adam optimizer.
%The model is trained on the AudioCaps~\cite{} dataset using the Adam optimizer.
%The learning rate is warmed up from $0$ to $10^{-4}$ over the first $5$ epochs,
The learning rate is warmed up to $10^{-4}$ in the first $5$ epochs, then decayed by a factor of $0.1$ every $10$ epochs. 
%Training employs the Adam optimizer, and 
The Cross Entropy loss is applied with a label smoothing factor $0.1$.
%The CrossEntropyLoss is applied with label smoothing factor set to $0.1$. 
Models are evaluated on AudioCaps, employing evaluation metrics including  
BLEU~\cite{kishore2002bleu}, $\text{ROUGE}_\text{L}$~\cite{chin-yew2004rouge}, METEOR~\cite{lavie2007meteor}, CIDEr~\cite{vedantam2015cider}, and SPICE~\cite{anderson2016spice}.
% BLEU~\cite{kishore2002bleu}, $\text{ROUGE}_\text{L}$~\cite{chin-yew2004rouge}, METEOR~\cite{lavie2007meteor}, CIDEr~\cite{vedantam2015cider}, SPICE~\cite{anderson2016spice}.
The fusion models are built on ACT-M and trained with the same  strategy.

%~\cite{liu2021swin}
HTS-AT uses an audio encoder comprises $4$ transformer blocks containing $2$, $2$, $6$, and $2$ Swin Transformer blocks respectively, followed by a CNN with 527 output channels.
%followed by a token-semantic Convolutional Neural Network (CNN) layer for classification. 
Constrained by data accessibility, the model is trained on a subset of AudioSet~\cite{gemmeke2017audio} containing approximately 1.9 million samples.
Nevertheless, we keep a fair comparison between the baseline and fusion experiments by utilizing the same data.
The fusion model is fine-tuned from the baseline model for a limited number of iterations to seek optimal performance on SED while preserving the original AT performance.
The training adopts AdamW optimizer with a warm-up strategy: The learning rates are set to $10^{-3}$, with a scaling factor of $0.05$, $0.1$, and $0.2$ during the first $3$ epochs then halved every $10$ epochs until $0.05$.
%until returning to $0.05$.
%Due to constraints in data resources, our model was trained on a subset of AudioSet~\cite{}.
The AT and SED tasks are tested on the AudioSet and DESED~\cite{serizel2020sound} datasets, respectively.
Evaluation metrics include mean average precision (mAP) for AT and event-based F1-score for SED.

\subsection{Results} 
The experimental results are presented in Tables~\ref{tab:audiocaps} and ~\ref{tab:audioset_desed}.
For the audio captioning task, the mid-fusion of features from generative models led to substantial improvements across all evaluation metrics. 
It outperforms the larger ACT-L with fewer trainable parameters, demonstrating the effectiveness of incorporating generative features.
Ablation studies are conducted to validate our hypotheses: (1) Simply replacing original features with generative ones resulted in performance degradation, which can be attributed to the insufficient semantic richness of the generative features alone. 
(2) Early fusion yielded modest performance gains, suggesting that the supplementary information provided by generative features is also discarded during progressive abstraction in audio encoder.
That the mid-fusion approach achieved optimal performance validates:  \textbf{The generative and audio encoder's discriminative  features  are indeed complementary}.

Results on SED and AT further corroborate our previous observations. 
% With the task granularity transitioning from fine to coarse—specifically from AAC to SED to AT—the performance gain is observed to progressively diminish.
%The performance improvement diminishes progressively as the task granularity becomes coarser—from AAC to SED to AT. 
The performance benefits offered by generative features gradually diminish from fine- to coarse-grained tasks (AAC$\rightarrow$SED$\rightarrow$AT), validating that \textbf{generative features primarily provide fine-grained information}.
%Ablation studies on these two tasks also confirm hypothesis 1.
% Compared to AAC, the improvement in SED—which demands less fine-grained information—was limited. 
% For the AT, which only requires classification, the fine-grained information from generative features contributed minimally to performance improvement.

These findings provide a new perspective on generative models and their features. 
Since generative models learn to approximate the true distribution of sounds in the real world, they capture fine-grained information that is complementary to high-level semantic features. 
In our view, the primary significance of these findings lies not in the mere enhancement of audio understanding tasks, but in the conceptual insights they contribute to the field of audio representation learning.
% This insight represents a valuable conceptual contribution of this work.
% This observation aligns with our hypothesis that fine-grained spatiotemporal features—inherent to generative representations—exert a more substantial influence on tasks requiring higher temporal or semantic resolution.

\section{Conclusion}
%In this work, we 
This work presents the first exploration of leveraging audio generative features for audio understanding tasks.
Discriminative models extract high-level semantic information through deep network layers, a process that often omits fine-grained acoustic details.
In contrast, generative models must generate high-fidelity audio with rich detail, enabling them to develop a robust perception of spatiotemporal acoustic properties.
%On the other hand, generative models are required to produce high-quality audio with rich detail, thereby developing a stronger perception of the spatiotemporal characteristics of acoustic events. 
We therefore propose to leverage generative features to complement the lack of spatiotemporal  information in discriminative features.
We validate our hypothesis across $3$ downstream tasks with varying granularity demands. 
The results show that generative features yield the most substantial gains in the fine-grained task of audio captioning, while offering minimal improvement in the coarse-grained task of audio tagging—empirically confirming that generative features encode detailed acoustic information.
% The results demonstrate that generative features bring the most significant improvement in the fine-grained task of audio captioning, while having limited impact on the coarse-grained task of audio tagging—providing empirical evidence that generative features indeed preserve acoustically detailed information. 
%Ablation studies further indicate that generative features alone lack sufficient semantic richness.
% , and early fusion leads to their informational contribution being obscured by subsequent layers of the audio encoder. 
% Thus, identifying an appropriate fusion strategy is critical; we show that mid-fusion serves as an effective approach. 
%We hope this study inspires further discussion on leveraging generative features for audio understanding tasks.
Beyond mere performance improvement, this work offers a novel perspective on audio processing, and we hope it can inspire new directions for audio representation learning.

\bibliographystyle{IEEEbib}
\bibliography{strings,refs}

\end{document}